\documentclass{mn2e}
\usepackage{graphicx,natbib2,natbibmnfix}
\usepackage{amsmath}
\usepackage{mathrsfs}
\usepackage{subfigure}
\usepackage{color}
\usepackage{url}
\usepackage{multicol}
\usepackage[table]{xcolor}

\newcommand{\be}{\begin{equation}}
\newcommand{\ee}{\end{equation}}

\def\ltsima{$\; \buildrel < \over \sim \;$}
\def\simlt{\lower.5ex\hbox{\ltsima}}
\def\gtsima{$\; \buildrel > \over \sim \;$}
\def\simgt{\lower.5ex\hbox{\gtsima}}

\def\del#1{{}}

\def\msun{{\,{\rm M}_\odot}}

\def\dx{x_{ij}}
\def\dy{y_{ij}}
\def\dz{z_{ij}}

\def\NovA{{\sc NovA}}
\def\Ramses{{\sc RAMSES}}
\def\Gadget{{\sc Gadget}}

\def\Power{{\sc Power}}
\def\Dehnen{{\sc Dehnen}}
\def\Romeo{{\sc Romeo}}

\def\Hahn{Valinia}

\title[{{\sc Nov}}el {{\sc A}}daptive softening for collisionless $N$-body simulations: \NovA]
{{{\sc Nov}}el {{\sc A}}daptive softening for collisionless $N$-body simulations: Eliminating spurious halos}

\author[Alexander Hobbs, Justin I. Read, Oscar Agertz, Francesca Iannuzzi, Chris Power]
       {\parbox{18cm}{Alexander Hobbs$^{1}$, Justin I. Read$^{2}$, Oscar Agertz$^{2}$, Francesca Iannuzzi$^{3}$, \\
       Chris Power$^{4}$}\vspace{0.3cm}\\
\noindent
$^{1}$Institute for Astronomy, ETH Z\"urich, Switzerland
$^{2}$Department of Physics, University of Surrey, Guildford, GU2 7XH, Surrey, UK\\
$^{3}$Laboratoire d'astrophysique de Marseille\\
$^{4}$University of Western Australia}

\begin{document}

\maketitle

\begin{abstract}
We describe a {{\sc Nov}}el form of {{\sc A}}daptive softening (\NovA) for collisionless $N$-body simulations, implemented in the \Ramses\ adaptive mesh refinement code. In \Ramses\ -- that we refer to as a `standard $N$-body method' -- cells are only split if they contain more than eight particles (a mass refinement criterion). Here, we introduce an additional criterion that the particle distribution within each cell be sufficiently isotropic, as measured by the ratio of the maximum to minimum eigenvalues of its moment of inertia tensor: $\eta = \lambda_{\rm max}/\lambda_{\rm min}$. In this way, collapse is only refined if it occurs along all three axes, ensuring that the softening $\epsilon$ is always of order twice the largest inter-particle spacing in a cell. This more conservative force softening criterion is designed to minimise spurious two-body effects, while maintaining high force resolution in collapsed regions of the flow. 

We test \NovA\ using an antisymmetric perturbed plane wave collapse (`\Hahn' test) before applying it to warm dark matter (WDM) simulations. For the \Hahn\ test, we show that -- unlike the standard $N$-body method -- \NovA\ produces no numerical fragmentation while still being able to correctly capture fine caustics and shells around the collapsing regions. For the WDM simulations, we find that \NovA\ converges significantly more rapidly than standard $N$-body, producing little or no spurious halos on small scales. We show, however, that determining whether or not halos exist below the free streaming mass $M_{\rm fs}$ is complicated by the fact that our halo finder (AHF) likely incorrectly labels some caustics and criss-crossing filaments as halos, while one or two particularly massive filaments appear to fragment in any version of \NovA\ where refinement is allowed. Such massive filaments may be physically unstable to collapse, as is the case for infinite, static, self-gravitating cylinders. We will use \NovA\ in forthcoming papers to study the issue of halo formation below $M_{\rm fs}$; filament stability; and to obtain new constraints on the temperature of dark matter. 
\end{abstract}

\renewcommand{\thefootnote}{\fnsymbol{footnote}}
\footnotetext[1]{E-mail: {\tt ahobbs@phys.ethz.ch}}

\section{Introduction}\label{sec:introduction}

The $N$-body method is widely used for modelling the non-linear growth of structure in the Universe \citep[e.g.][]{2011EPJP..126...55D,2012PDU.....1...50K}. For collisionless non-relativistic (`cold') dark matter, it has been shown to be remarkably accurate, producing robust results that are numerically well-converged across different implementations \citep[e.g.][]{2007arXiv0706.1270H,2009MNRAS.398L..21S,2008MNRAS.391.1685S,2013arXiv1308.2669K}. Such simulations provide an excellent match to the observed large scale structure in the Universe \cite[e.g.][]{2006Natur.440.1137S}, though on smaller scales -- where baryons likely play a role \citep[e.g.][]{NavarroEkeFrenk1996,ReadGilmore2005,MashchenkoEtal2008,2011arXiv1106.0499P} -- there are known discrepancies \citep[e.g.][]{1994ApJ...427L...1F,1994Natur.370..629M,1999ApJ...524L..19M,1999ApJ...522...82K}.

Despite the successes of the $N$-body method, since its inception there have been concerns about the effect of discreteness errors on numerical accuracy and convergence \citep[e.g.][]{1998ApJ...497...38S,1997ApJ...479L..79M,2004MNRAS.348..977D,2004MNRAS.350..939B,2007MNRAS.380...93W,2008ApJ...686....1R,2009MNRAS.394..751J}. These arise because the dark matter fluid is represented by a set of `particles', each with mass typically in the range $10^3 - 10^6$\,$M_\odot$. To avoid spurious scattering between these particles the force is softened, for example using `Plummer' \citep{1915MNRAS..76..107P} softening: 

\begin{equation}
{\bf F}_{ij} = \frac{Gm^2({\bf x}_j-{\bf x}_i)}{(\epsilon^2 + |{\bf x}_i-{\bf x}_j|^2)^{3/2}}
\label{eqn:forceij}
\end{equation}
where ${\bf F}_{ij}$ is the force between two particles $i$ and $j$ at positions ${\bf x}_{i,j}$; $G$ is Newton's gravitational constant; and $\epsilon$ is the force softening. Equation \ref{eqn:forceij} ensures that the force is clipped at a constant value as two particles approach one another. This reduces spurious two-body scattering, but it does not prevent numerical relaxation from occurring; that can only be combated by raising the number of particles \citep[e.g.][]{2001MNRAS.324..273D,2003MNRAS.338...14P,2004MNRAS.350..939B,2004MNRAS.348..977D,2011EPJP..126...55D}. 

For cold dark matter (CDM) simulations, numerical convergence appears to be very good \citep[e.g.][]{2007arXiv0706.1270H}. However, discreteness errors may yet play a role when attempting to calculate power spectra, mass functions and higher order halo statistics at percent level accuracy, as will be required by next generation cosmological probes \citep[e.g.][]{2013MNRAS.431.1866R,2013LRR....16....6A}. More problematic, however, are simulations that model a sharp cut-off in the initial power spectrum, as in {\it warm} dark matter\footnotemark (WDM;   \citealt{2001ApJ...556...93B,2001ApJ...559..516A}), or exotic inflationary models \citep{2003ApJ...598...49Z}. The first WDM simulations appeared to find evidence of fragmentation -- smaller halos forming later than larger ones -- as evidenced by a sharp upturn in the halo mass function \citep[e.g.][]{2001ApJ...556...93B,2001ApJ...559..516A}. However, this has now been traced to the numerical fragmentation of filaments due to discreteness effects. This is particularly worrisome since the `spurious halos' that form via this process diminish with particle number only as $N^{1/3}$, leading to extremely slow convergence \citep{2007MNRAS.380...93W}.

\footnotetext{In WDM, it is supposed that the dark matter is non-relativistic for a time after decoupling, leading to a suppression in the growth of structure on small scales and at early times \citep[e.g.][]{2001ApJ...556...93B,2001ApJ...559..516A}. Typically, this is modelled as an exponential cut-off in the initial power spectrum and indeed throughout this paper, where we refer to WDM simulations, this is what we mean. Fully self-consistent WDM models (for example, sterile neutrinos) have more complex model-dependent power spectra than this \citep[e.g.][]{2009ARNPS..59..191B}. Furthermore, for hot dark matter, it can also become important to model the primordial velocity dispersion of the dark matter particles. This has been attempted only a few times in the literature, most likely because of the computational cost involved. A proper treatment requires us to replace each `cold dark matter' particle in the initial conditions by $\sim 1000 - 10,000$ lighter particles in order to well-sample the local velocity distribution function at each point in the flow \citep[e.g.][]{2001ApJ...559..516A,2013MNRAS.434.1171H}. As far as the authors are aware, at the time of writing, such an expensive approach has never been attempted.} 

The likely reason for the formation of `spurious halos' in WDM simulations was only recently elucidated by \citet{2013MNRAS.434.1171H}. Using a new method for evolving collisionless fluids -- where they track the foliations of the the 3D dark matter phase sheet -- they find that the spurious halos result from large anisotropic force errors. With a more accurate force (as calculated by their new method), the spurious halos are much reduced, and the resulting filaments are smooth.

While \citet{2013MNRAS.434.1171H} present an elegant alternative to $N$-body simulations, their method is numerically expensive. Since they are required to track the folding of the phase sheet, at the centres of dark matter halos where there are many such foliations they formally require an ever-increasing number of simulation elements \citep{2015arXiv150101959H}; without such refinement, unphysical behaviour occurs in high density regions. By contrast, a key strength of the `standard' $N$-body method is that, since the equations of motion are derived from a Hamiltonian, the time-averaged expectation value of the energy of a particle will be correct even if its orbital phase is wrong\footnote{This is only strictly true if a symplectic time integrator is used. The Leapfrog integrator typically employed in cosmological simulations is symplectic, but only for fixed timesteps \citep[e.g.][]{2011EPJP..126...55D}.} \citep[e.g.][]{2011EPJP..126...55D}. This likely explains the success of the $N$-body method even at rather modest $N$. For example, \citet{2006astro.ph.10468S} find convergence for their disc simulations already with $N \sim 10^5$, despite earlier calculations suggesting some $\sim 10^8$ particles would be required to properly resolve resonances \citep{2007MNRAS.375..425W}.

The above motivates considering whether the classic $N$-body method cannot be improved. Two recent works have attempted to `repair' $N$-body simulations in post-processing by pruning spurious halos. \citet{2013MNRAS.433.1573S} propose fitting a power law to the artificial halos and subtracting them away, taking advantage of the fact that spurious halos are more prevalent in over-dense regions. By contrast, \citet{2014MNRAS.439..300L} suggest an algorithm where subhalos are removed from the mass function if: (i) their `protohalos' are highly flattened; and/or (ii) the subhalos are below a mass cut; and/or (iii) the subhalos are not present in a higher resolution simulation of the same halo. In this paper, we consider instead a modified {\it force softening} criterion. This is designed to improve the anisotropic force errors that are at the root of the problem, leading to a more faithful $N$-body method in the first place.

This paper is organised as follows. In \S\ref{sec:softening}, we briefly review different strategies in the literature for force softening and we present our new {{\sc Nov}}el form of {{\sc A}}daptive softening -- \NovA\ -- designed to minimise spurious two-body effects. In \S\ref{sec:numerics}, we describe our \NovA\ algorithm in detail and its implementation in the \Ramses\ code. In \S\ref{sec:results}, we compare \NovA\ to standard \Ramses\ for an asymmetric plane wave test and 0.2\,keV WDM simulations. We focus here on presenting the first results from \NovA\ for the density field; mass function; and dark matter halo density profiles. A detailed analysis of halo formation below the WDM `free-streaming' mass (see \S\ref{sec:wdmics}); filament stability; and obtaining new constraints on the temperature of dark matter using \NovA\ will be presented in forthcoming publications. Finally, in \S\ref{sec:conclusions} we present our conclusions.

\section{Force softening}\label{sec:softening}

Since spurious halos in WDM simulations appear to result from anisotropic force errors, this suggests that a good place to begin in improving the $N$-body method is to take a critical look at how the force softening $\epsilon$ is chosen\footnote{Note that it is equally important to select an appropriate timestep for the particles \citep[e.g.][]{2003MNRAS.338...14P,2007MNRAS.376..273Z}. However, this is true both in the standard $N$-body method and in the `folding phase sheet' model of \citet{2013MNRAS.434.1171H}. This suggests that either the choice of $\epsilon$ is more crucial than the choice of timestep, or that the timestep criteria typically used in $N$-body simulations \citep[e.g.][]{2011EPJP..126...55D} are adequate.}. Ideally, $\epsilon$ should be as small as possible such that the maximum possible force resolution is obtained for a given numerical cost. However, too small and spurious forces will creep in, potentially spoiling numerical convergence. We consider three different force softening strategies here: 

\begin{enumerate}

\item {\it Minimising two-body effects} (\Power): A popular rule-of-thumb, that has been carefully tested on cold dark matter (CDM) simulations, follows from ensuring that two body forces are small as compared to the mean field \citep{2003MNRAS.338...14P}: 

\begin{equation} 
\frac{G m^2}{\epsilon^2} \sim \frac{1}{\alpha^2} \frac{G M m}{r^2} \Rightarrow \epsilon \sim \alpha \frac{r}{\sqrt{N}}
\label{eqn:powersoft}
\end{equation}
where $\alpha = 4$ is an empirically derived parameter \citep{2003MNRAS.338...14P}; $N = M/m$ is the number of particles inside some characteristic radius $r$ (\citealt{2003MNRAS.338...14P} use the virial radius $r_{200}$); and $m$ is the dark matter particle mass. \\

\item {\it Minimising force errors} (\Dehnen): An alternative approach is to minimise errors coming from biased forces that occur if $\epsilon$ is too large, and noise that occurs if $\epsilon$ is too small \citep{2001MNRAS.324..273D}. This leads to a well-defined optimal force softening that depends on the particular gravitational potential being simulated (and the choice of softening kernel). For small $\epsilon$ and large $N$, \citet{2001MNRAS.324..273D} derive an analytic estimate for Plummer softening that scales as:

\begin{equation} 
\epsilon \propto N^{-0.73}; \,\,\,\, {\rm valid\,\,for\,\,} \epsilon \ll r; N \simgt 10^5
\label{eqn:dehnensoft}
\end{equation}

\item {\it Minimising scatter between an ensemble of $N$-body realisations} (\Romeo): Finally, following earlier work by \citet{1997ApJ...479L..79M} and \citet{1998ApJ...497...38S}, \citet{2008ApJ...686....1R} take a different approach. They run ensembles of the same cosmological $N$-body simulation, varying only the random number seed and the force softening. They argue that the scatter in results (as measured by various metrics like the halo mass function) has a term that is physical (cosmic variance), and a term that is numerical (discreteness noise). The optimum force softening should minimise the discreteness noise and therefore should minimise the scatter between different realisations of the same simulation. Using a novel wavelet analysis, they empirically derive:

\begin{equation}
\epsilon \sim 2d
\label{eqn:romeosoft}
\end{equation} 
where $d$ is the mean inter-particle spacing.

\end{enumerate} 
While each of the above approaches to force softening is conceptually different, they all point to a rather similar conclusion: {\it $\epsilon$ must be adaptive, varying both in space and time}: $\epsilon \equiv \epsilon({\bf x},t)$. To see this, it is instructive to consider a simple toy model where the dark matter is spherically-distributed with a power-law density profile:

\begin{equation} 
\rho \propto r^{-\gamma} \Rightarrow N(<r) \propto r^{(3-\gamma)}
\label{eqn:rhosimp}
\end{equation}
where $N(<r)$ is the cumulative number of particles within $r$. (This equation is only strictly valid near the centre of dark matter halos.) For this toy model, our three criteria give rather different results, but all point towards $\epsilon$ being some function of radius $r$ and therefore of the local density: 

\begin{equation}
\epsilon \propto \rho^{-\kappa}; \,\,\,\, \kappa > 0 
\label{eqn:kappasoft}
\end{equation}
where: 
\begin{equation}
\kappa = \left\{
\begin{array}{ll} 
(\gamma-1)/2\gamma & {\rm {\sc \Power}} \\
0.73(\gamma-3)/\gamma & {\rm {\sc \Dehnen}} \\ 
1/3 & {\rm {\sc \Romeo}} \\
\end{array} 
\right.
\nonumber
\end{equation} 
There are several interesting points to note from equation \ref{eqn:kappasoft}. Firstly, notice that the \Power\ criterion actually amounts to {\it fixed softening} if $\gamma = 1$, as is the case for the centres of CDM halos \citep{1991ApJ...378..496D}. This may explain why fixed softening simulations have performed so surprisingly well despite the natural expectation that $\epsilon$ ought to be adaptive. If we are to adapt $\epsilon$, however, the \Power\ criterion becomes potentially pathological. For $\gamma < 1$, $\kappa < 0$ and in shallow dark matter cusps or cores, the softening would actually {\it increase} with density. The \Dehnen\ criterion fares better in this regard, being well behaved for all $\gamma < 3$; however, more work is required to generalise it to larger radii where the softening will be large and equation \ref{eqn:dehnensoft} is then no longer valid. For these reasons, we consider only the \Romeo\ criterion from here on. 

The \Romeo\ softening has (perhaps inadvertently) been extensively explored in the literature. Mesh-based methods like the \Ramses\ code \citep{Teyssier2002} tie the softening to the local cell size which is naturally adaptive. Cells are split if they have greater than $N_{\rm c}$ particles, typically chosen to be $N_c \sim 8$ to achieve, on average, one particle per cell after cell refinement. This amounts to a scheme where $\epsilon \propto 1/\rho_{\rm local}^{1/3}$, exactly as in the \Romeo\ force softening. Similar schemes have also been explored in Tree $N$-body codes. There, since the equations of motion are derived from a Hamiltonian, it is possible to craft a density-adaptive $\epsilon$ method that is manifestly conservative \citep{Price2007}. \citet{2011MNRAS.417.2846I} have recently implemented this in the \Gadget\ code \citep{2005MNRAS.364.1105S}, finding that it leads to results in excellent agreement with the fixed $\epsilon$ case, while giving greater resolution for similar numerical cost (see also \citealt{2013MNRAS.428.1968K}). Their results suggest that with or without the conservative correction terms, the halo mass function converges; however, without the correction there is a substantial suppression of low-mass halos mass as compared to both the conservation-corrected and fixed softening simulations. In Appendix \ref{conservative-soft}, we show that such conservative corrections do not solve `spurious halo' problem in WDM simulations. In fact, since the correction terms appear as a purely attractive force that points along the density gradient, they make the spurious halo problem worse.

\begin{figure}
\includegraphics[width=0.49\textwidth]{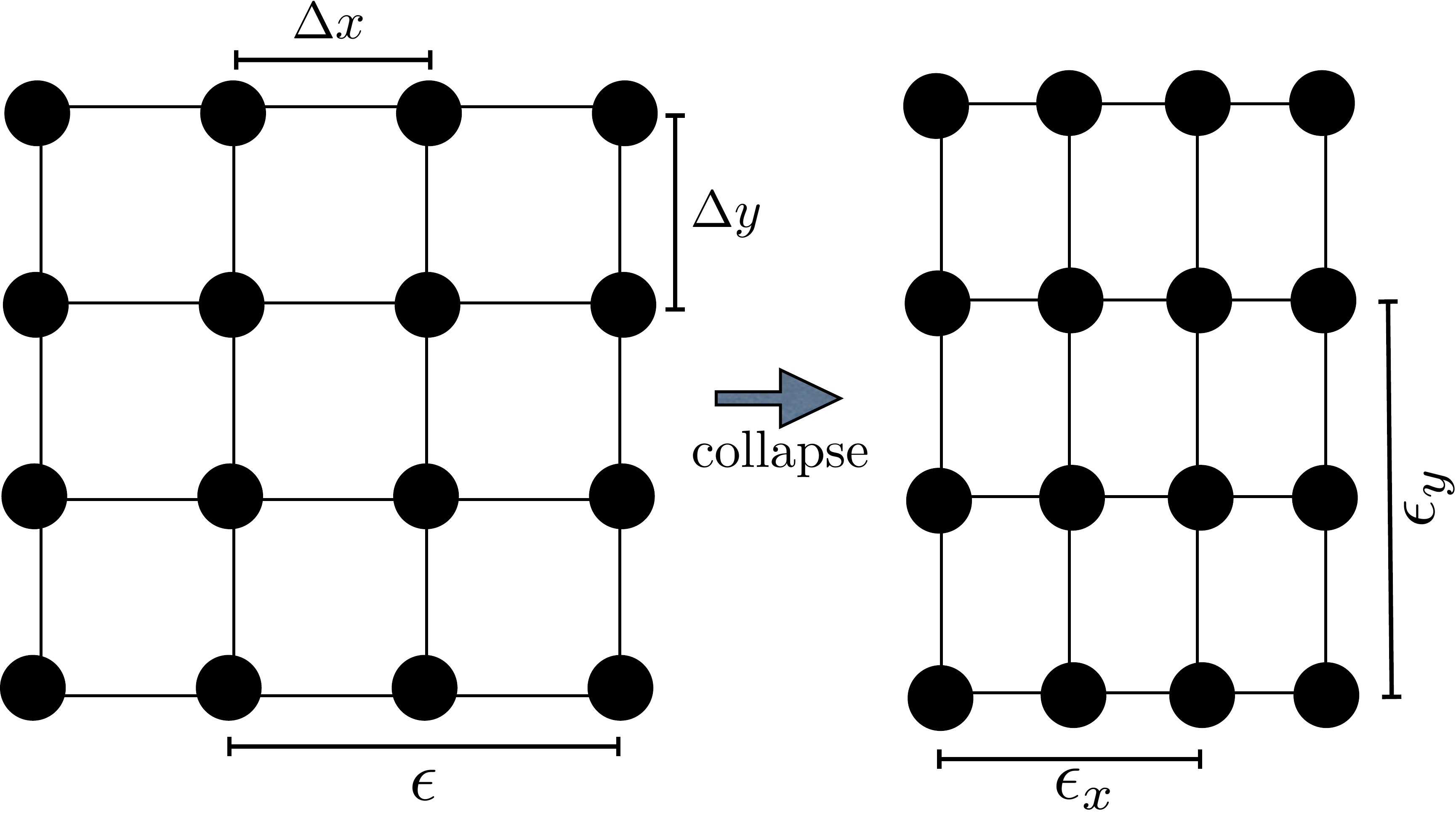}
\caption{A schematic view of the need for a modified adaptive force softening for $N$-body simulations. The simulation begins with the distribution locally very close to isotropic (left), with mean interparticle spacing $\Delta x \sim \Delta y$. However, as collapse proceeds first along the shortest axis (in this case the $x$ axis), we quickly move to a situation where $\Delta x \ll \Delta y$ (right). Standard adaptive softening schemes adapt purely on the local density. In this case, we would pick a softening $\epsilon \propto \Delta x \equiv \epsilon_x$, making the softening too small in the $y$ direction. This could lead to spurious clumping along the filament. In \NovA\, we measure the local anisotropy and do not refine $\epsilon$ if the anisotropy is too high. In the example pictured, this would set our softening proportional to the longest local axis of the collapse -- in this case $\epsilon_y \propto \Delta y$. The softening remains isotropic, but is more conservative than standard schemes in regions of high anisotropy.}
\label{fig:anisotropycartoon}
\end{figure}

In this paper, we present a {{\sc Nov}}el form of {{\sc A}}daptive softening -- \NovA\ -- designed to minimise spurious two-body effects. Like the \Romeo\ softening, we tie the softening to the local interparticle spacing $\epsilon \sim 2d$. However, for the first time we account for the fact that in cosmological simulations, collapse is expected to be locally {\it anisotropic} \citep[e.g.][]{1978IAUS...79..409Z}. Since collapse proceeds most rapidly along the short axis of the flow, in the early stages of collapse there will always be {\it three} interparticle spacings aligned along the short $c$, intermediate $b$, and long $a$ axis (see Figure \ref{fig:anisotropycartoon}). If we adapt $\epsilon$ purely on density, this amounts to an optimistic criterion $\epsilon \sim 2c$ that actually violates the \Romeo\ criterion along the long axis, leading to potentially large spurious two-body forces. Instead, our new \NovA\ method allows $\epsilon$ to be adapted on density \emph{only if} the collapse is sufficiently isotropic. We show that this simple change to the $N$-body algorithm prevents the formation of `spurious' halos along filaments. (Note that, while in this paper we adapt on density, in principle \NovA\ can be applied to any adaptive softening scheme that obeys equation \ref{eqn:kappasoft}. We defer such generalisations to future work.)

\section{Numerics}\label{sec:numerics}

\subsection{The \Ramses\ `standard $N$-body' code}

We carry out cosmological $N$-body simulations using the Adaptive Mesh Refinement (AMR) code \Ramses\ \citep{Teyssier2002}. The collisionless dark matter dynamics are evolved using the particle-mesh technique \citep[see e.g.][]{1988csup.book.....H}, with gravitational accelerations computed from the gravitational potential on the mesh. The gravitational potential is calculated by solving the Poisson equation using the multi-grid method \citep{2011JCoPh.230.4756G} for all refinement levels.  
 
Note that while we describe \Ramses\ as a `standard $N$-body' code, it actually differs from most Tree $N$-body codes in an important respect: the softening $\epsilon$ is automatically adapted according to the \Romeo\ criterion if the refinement strategy is based on reaching a critical number of particles per cell. Since this is true for all adaptive mesh refinement schemes in the literature to date, however, we still refer to this as `standard'. We compare and contrast \Ramses\ with some \Gadget\ Tree $N$-body simulations that use both fixed and adaptive softening in Appendix \ref{conservative-soft}.

\subsection{The \NovA\ algorithm} 

In this section, we describe our {\sc Nov}el {\sc A}daptive force softening algorithm: \NovA. This is a modified cell splitting criterion implemented in the \Ramses\ code. Normally, cells are split if they contain more than some critical number of particles in a cell: $N_{\rm cell} > N_c$. Here, we add an additional criterion that the cell is sufficiently isotropic as measured by its moment of inertia tensor: 

\begin{equation}
I_i = \sum\limits_{j=1}^{N} \left[ \begin{array}{ccc}
(\dy^2 + \dz^2) \; m_j  & - \dx \; \dy \; m_j & - \dx \; \dz \; m_j \\
- \dx \; \dy \; m_j & (\dx^2 + \dz^2) \; m_j & - \dy \; \dz \; m_j \\
- \dx \; \dz \; m_j & - \dy \; \dz \; m_j & (\dx^2 + \dy^2) \; m_j \end{array} \right]
\end{equation}
where $\dx = x_i - x_j$, $\dy = y_i - y_j$, $\dz = z_i - z_j$. 

We compute the eigenvalues of the matrix $I_i$: $\lambda_1, \lambda_2, \lambda_3$, which are sorted so that $\lambda_1 > \lambda_2 > \lambda_3$, and take the ratio $q_{\rm i} \equiv \lambda_1/\lambda_3$ to be a measure of the (spatial) anisotropy in the particle distribution.

Cells are split if $N_{\rm cell} > N_c$ {\it and} $q_i < \eta$, where $\eta$ is a parameter that controls the amount of anisotropy allowed for splitting to occur. The effect of this is shown schematically in Figure \ref{fig:anisotropycartoon}. By refining only where the particle distribution is locally isotropic, \NovA\ effectively picks the most conservative local force softening. As a result, the softening in anisotropic regions is always at least twice the {\it longest} inter-particle spacing in a cell: $\epsilon \simgt 2 \max[d]$. This means that the softening is somewhat overestimated with respect to the short axis. However, too large softening only affects the computational efficiency (since for the same particle number we have reduced force resolution); whereas too small softening can -- through two body effects -- be much more problematic. An alternative way to think of the algorithm is that it does not refine unless collapse is occurring along all three axes. This typically reduces refinement in filamentary or elongated structures.


\subsection{The choice of $\eta$ and $N_c$}

In the limit $N \rightarrow \infty$, we would ideally have an anisotropy bound of $\eta = 1$ -- i.e. cell splitting is allowed only for purely isotropic cells. However, in practice noise in the particle distribution makes it undesirable to set $\eta = 1$ exactly. Here, we choose as default $N_c = 32$ which is chosen to ensure that there are always enough particles in a cell that $I_i$ can be reliably measured ($N_c = 32$ ensures at least $\sim 3$ particles per spatial dimension); and we select $\eta = 1.08$. The latter number is chosen by drawing 32 particles at random from a uniform density distribution and calculating the distribution function of $q$. We choose $\eta  = 1 + 0.5 \sigma_{q}$ where $\sigma_{q}$ is the variance in $q$ for this random sampling; similar results for a range of $N_c$ are reported in Table \ref{tab:etancell}. This has two desirable properties: (i) $\eta$ is set by the noise level for a cell; (ii) as a result, $\eta$ naturally shrinks with $N_c$. Note that the above implies that varying $\eta$ with the number of particles in a cell, or with the refinement level may give improved performance over the fixed $\eta$ scheme we explore here; such improvements are beyond the scope of this present work. We explore the effect of varying $N_{\rm cell}$ and $\eta$ in \S\ref{sec:etancell}.

\begin{table}
\begin{center}
\begin{tabular}{lll}
$N_c$ & $\sigma_q$ & $\eta$ \\
\hline
8  &                 0.748626 & 1.374\\
16  &                0.304557 & 1.152\\
\hline
\rowcolor{gray!50}
32 &                 0.169258 & 1.084\\
\hline
64  &                0.102809  & 1.051\\
128 &             0.0661424 & 1.033\\
256   &            0.0448105 & 1.022\\
512 &              0.0301317 & 1.015\\
\end{tabular}
\end{center}
\caption{Using cell particle noise to choose the anisotropy parameter $\eta$ for a given number of particles per cell $N_c$. The columns show $N_c$; the variance in anisotropy parameter for random realisations of a uniform particle distribution $\sigma_q$; and our choice of $\eta$ derived from this analysis: $\eta = 1 + 0.5 \sigma_q$. Our approach has two desirable properties: (i) $\eta$ is set by the noise level for a cell; (ii) as a result, $\eta$ naturally shrinks with $N_c$. The grey row marks our default choice of parameters.}
\label{tab:etancell}
\end{table}

\subsection{Numerical performance}

As with any new numerical algorithm, we would be remiss not to discuss its numerical cost. We find that the relative performance of \NovA\ to \Ramses\ depends on the precise choice of parameters and problem setup. Our default \NovA\ scheme is $\simgt 10$ times faster than \Ramses\ with $N_c = 8$. However, this is simply because \NovA\ refines less. When comparing with standard \Ramses\ with $N_c = 32$ (that refines only one level deeper than \NovA; see Table \ref{tab:tableics}), \NovA\ is almost the same speed. If we compare \NovA\ and \Ramses\ for simulations where \NovA\ refines similarly to \Ramses, \NovA\ is about $\sim 20\%$ slower.

\section{Results}\label{sec:results}

\subsection{The asymmetric plane wave (\Hahn) test}\label{sec:hahntest}

\begin{figure}
\includegraphics[width=0.49\textwidth]{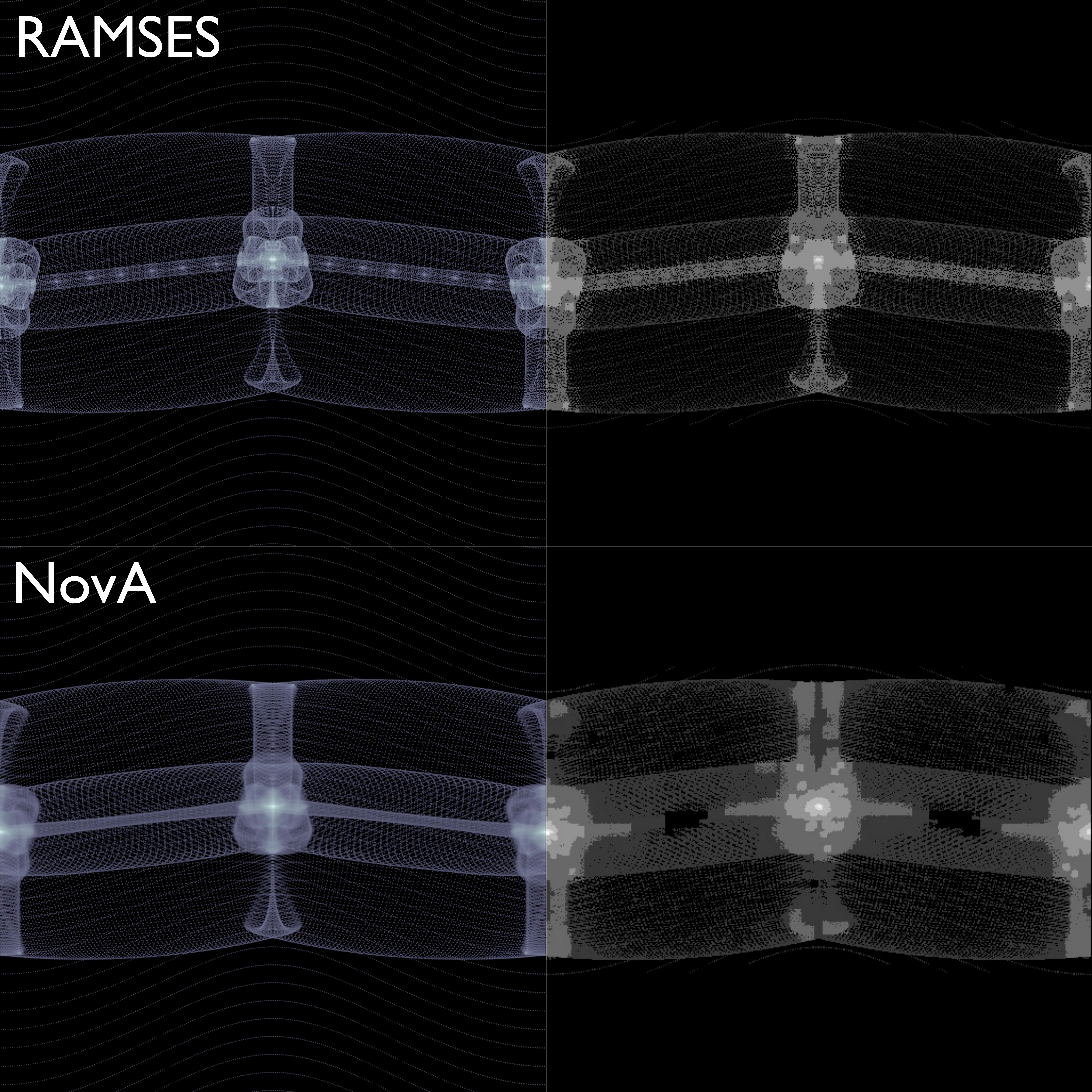}
\caption{$N = 512^3$ run for the asymmetrically-perturbed plane wave (\Hahn) test, comparing standard \Ramses\ (top) and our default \NovA\ method (bottom). The left panels show the particle distribution; the right the AMR refinement map. Notice that in \Ramses, the filaments break up into regularly-spaced clumps. This occurs because the standard cell-splitting criterion refines on the filaments (see top right panel). By contrast, in \NovA\ no such refinement occurs (see bottom right panel) and the filaments are smooth. Both algorithms refine on the bound structures that form at nodes, capturing the same caustic/shell-like structures reported in \citet{2013MNRAS.434.1171H}.}
\label{fig:hahn_composite}
\end{figure}

We set up initial conditions for an asymmetric plane wave test as in \citet{1997ApJ...479...46V} and \citet{2013MNRAS.434.1171H}. This simple 2D test allows us to make a check on two-body discreteness effects without running a full cosmological volume. The plane wave is setup along the x-direction with the following (sinusoidal) phase perturbation in the y-direction:

\begin{equation}
\phi(\vec{x}) = \bar{\phi} \cos \left( k_{\rm p} \left[ x + \epsilon_{\rm a} \frac{k_{\rm p}}{k_{\rm a}^2} \cos k_{\rm a} y \right] \right)
\end{equation}
where $k_{\rm p} = 2 \pi/L$, $k_{\rm a} = 4 \pi/L$ and $\epsilon_{\rm a} = 0.2$. $L$ refers to the size of the simulation box. $\bar{\phi}$ sets the value of expansion factorat which the first shell crossing occurs -- this is set to $a_{\rm c} = 1/7.7 \simeq 0.13$. The initial particle positions and velocities were obtained by applying the Zel'dovich approximation \citep{Zeldovich1970} to an unperturbed regular Cartesian lattice.

The plane wave is allowed to evolve under the pure gravitational potential of the particles up until $a = 1$. The results for our default choice of $N_c = 32$ and $\eta = 1.08$ are shown in Figure \ref{fig:hahn_composite}. The left panels show the particle distribution; the right the AMR refinement map. Notice that in \Ramses, the filaments break up into regularly spaced clumps. This occurs because the standard cell splitting criterion refines on the filaments (see top right panel). By contrast, in \NovA\ no such refinement occurs (see bottom right panel) and the filaments are smooth. Both algorithms refine on the bound structures that form at nodes, capturing the same caustic/shell-like structures reported in \citet{2013MNRAS.434.1171H}. Since the bound lumps move to lower mass and smaller spacing with resolution (they are non-convergent), \NovA\ gives a more faithful simulation of the correct physics. \NovA\ -- unlike the standard \Ramses\ $N$-body implementation -- converges much more rapidly with resolution. We will show this more quantitively with warm dark matter (WDM) simulations, next.

\subsection{Warm dark matter simulations}\label{sec:wdmsims}

\subsubsection{Initial conditions and simulation analysis}\label{sec:wdmics}

The Warm Dark Matter (WDM) simulations were set up as in \citet{PowerEtal2003}. We used cosmological parameters $\Omega_{\rm 0} = 0.27$, $\Omega_{\rm \Lambda} = 0.73$, $h = 0.705$ and $\sigma_{\rm 8} = 0.81$ at $z = 0$ \citep{KomatsuEtal2011}. Initial conditions were created by generating a statistical realization of a Gaussian random density field in Fourier space, with variance given by the linear matter power spectrum, and the Zel'dovich approximation used to compute initial particle positions and velocities. A CDM power spectrum was obtained by convolving the primordial power spectrum $P(k) \propto k^{n_{\rm spec}}$ with the transfer function appropriate for our chosen set of cosmological parameters, computed using the Boltzmann code \rm{CAMB} \citep[see][]{LewisEtal2000}. The power spectrum for the WDM model was then obtained a la \citet{BodeEtal2001}, by filtering the CDM power spectrum with an additional transfer function of the form
\begin{equation}
T^{\rm WDM} (k) = \left(\frac{P^{\rm WDM} (k)}{P^{\rm CDM} (k)}\right)^{1/2} = \left[1 + (\alpha k)^{2\nu}\right]^{-5/\nu}
\end{equation}
where $\alpha$ is a function of the WDM particle mass (eq. 9 in \cite{BodeEtal2001}), $k$ is the wavenumber and $\nu=1.2$ is a numerical constant.

We chose a WDM thermal relic mass of $m_\chi = 0.2$\,keV. Following \citet{2012MNRAS.424..684S}, this corresponds to an effective `free-streaming' scale of: 

\begin{equation}
\lambda_{\rm fs}^{\rm eff} \simeq 0.049 \left(\frac{m_\chi}{\rm keV}\right)^{-1.11}\left(\frac{\Omega_\chi}{0.25}\right)^{0.11}\left(\frac{h}{0.7}\right)^{1.22}\,{\rm Mpc}/h
\end{equation}
which for $\Omega_\chi = 0.25$; $h=0.73$; and $m_\chi = 0.2$ gives $\lambda_{\rm fs}^{\rm eff} = 0.308\,{\rm Mpc}/h$. And a `free-streaming mass scale':

\begin{equation}
M_{\rm fs} = \frac{4\pi}{3} \overline{\rho} \left(\frac{\lambda_{\rm fs}^{\rm eff}}{2}\right)^3
\end{equation}
where $\overline{\rho}$ is the mean background density of the Universe (that is a function of cosmology and redshift $z$; $\overline{\rho(z=0)} = 277.3 h^2$\,M$_\odot$\,kpc$^{-3}$; \citealt{1999coph.book.....P}). For purely linear collapse with no mode-coupling or fragmentation, no halos should form below $M_{\rm fs}$. At reshift $z=0$, and assuming the above cosmological parameters and WDM thermal relic mass, we have $M_{\rm fs} = 2.26 \times 10^9$\,M$_\odot$. 

A second scale of interest is the length scale at which the WDM transfer function is reduced by half: the `half-mode' length:

\begin{equation}
\lambda_{\rm hm} \simeq 13.93 \lambda_{\rm fs}^{\rm eff}
\end{equation}
which also has an associated mass scale, the `half-mode mass': 

\begin{equation} 
M_{\rm hm} \simeq 2.7 \times 10^3\,{\rm M}_{\rm fs}
\end{equation}
This is the mass scale at which we expect the WDM mass function to noticeably deviate from the CDM case.

\begin{figure*}
\includegraphics[width=0.99\textwidth]{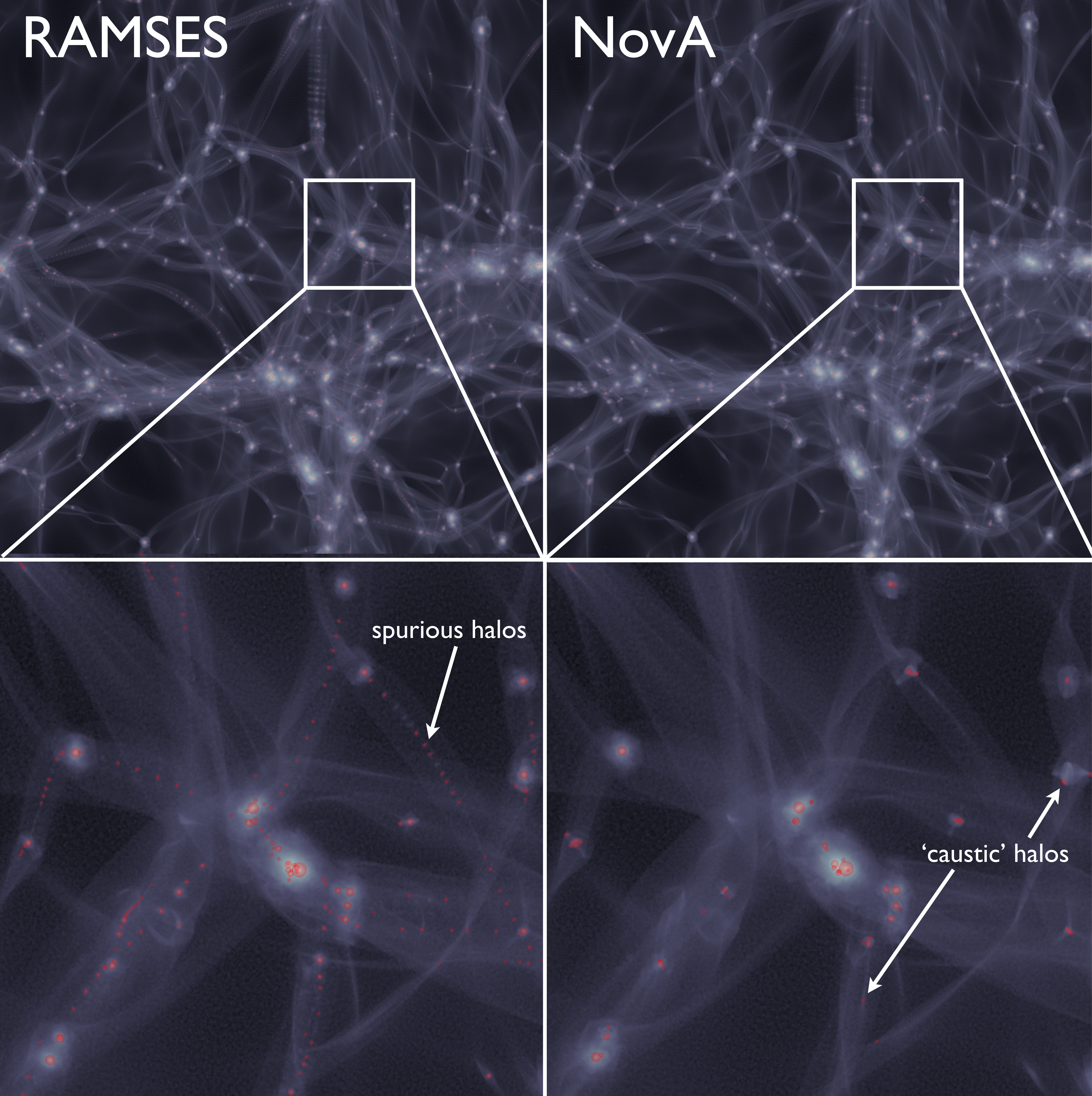}
\caption{A comparison of \Ramses\ and \NovA\ for a 0.2\,keV WDM simulation with $N = 512^3$ particles. The top panels show the full 50\,Mpc/h box; the bottom panels highlight a zoomed in region, as marked. Halos identified using AHF are marked by the red filled circles; their size is proportional to their virial radii. Notice that in \Ramses, many small and regularly spaced halos -- `spurious halos' -- form along the filaments; in \NovA\ these are gone. Notice further that in the zoom panel for \NovA, there are several cases of halos identified by AHF that may not correspond to genuine bound structures that could host galaxies. We highlight two of these `caustic' halos as examples. Some of these halos likely owe to overlapping caustics in the WDM density field and would not host galaxies; others, however, may be genuine non-linear structures that form at an overlap between caustics or as filaments intersect. This latter possibility is very interesting as it would imply that halos can form below the free streaming mass $M_{\rm fs}$ in WDM. We will explore this further in a forthcoming paper.}
\label{fig:density_composite}
\end{figure*}

It is not clear if halos should exist below $M_{\rm fs}$ in WDM structure formation. \citet{2013MNRAS.434.3337A} use their new `folding phase sheet' method to argue that there are no halos below $M_{\rm fs}$ (other than substructure halos that originate from halos more massive than $M_{\rm fs}$). However, this result relies on some manual pruning of the halo mass function -- required due to errors in the halo finding algorithm. We verify that halo finding in WDM is indeed a thorny issue, and discuss \NovA\ results for halos below $M_{\rm fs}$ in \S\ref{sec:denfield}.

\begin{table*}
\caption{The cosmological simulations and their parameters. The columns show from left to right: the simulation label; the particle resolution; the minimum number of particles per cell $N_c$; the anisotropy parameter $\eta$ (see \S\ref{sec:numerics}); the dark matter particle mass $m_{\rm part}$; and the maximum refinement level reached by that simulation.}
\centering
\begin{tabular}{|c|c|c|c|c|c|}\hline
\hline
Label & Resolution & $N_c$ & $\eta$ & $m_{\rm part}$ ($10^{10} \msun/h$) & Max. refinement level\\
\hline
RAMSES-256 & $256^3$ & $32$ & none & $7.1 \times 10^{-2}$ & 14\\
RAMSES-512 & $512^3$ & $32$ & none & $8.9 \times 10^{-3}$ & 16\\
RAMSES-1024 & $1024^3$ & $32$ & none & $1.1 \times 10^{-3}$ & 17\\
\hline
NovA-256 & $256^3$ & $32$ & $1.08$ & $7.1 \times 10^{-2}$ & 13\\
NovA-512 & $512^3$ & $32$ & $1.08$ & $8.9 \times 10^{-3}$ & 15\\
NovA-1024 & $1024^3$ & $32$ & $1.08$ & $1.1 \times 10^{-3}$ & 16\\
NovA-256-Nc128 & $256^3$ & $128$ & $1.03$ & $7.1 \times 10^{-2}$ & 7\\
NovA-512-Nc128 & $512^3$ & $128$ & $1.03$ & $8.9 \times 10^{-3}$ & 8\\
NovA-1024-Nc128 & $1024^3$ & $128$ & $1.03$ & $1.1 \times 10^{-3}$ & 14\\
\hline

\hline
\hline
\end{tabular}
\begin{flushleft}
\end{flushleft}
\label{tab:tableics}
\end{table*}

Note that we deliberately choose a small $m_\chi = 0.2$\,keV even though such a low thermal relic mass is already ruled out by constraints from the Lyman-$\alpha$ forest \citep[e.g.][]{2013PhRvD..88d3502V}. The reason for this is that it ensures that WDM effects will appear at large mass, making it computationally efficient to test our methodology \citep[c.f.][]{2013MNRAS.434.1171H}. We will present \NovA\ simulations of particle masses close to the current observational constraints (and comparisons with data) in future work. A full list of all simulations run and their parameters is given in Table \ref{tab:tableics}.

\subsubsection{The density field}\label{sec:denfield}

Figure \ref{fig:density_composite} shows a full box view of the projected dark matter density field in \Ramses\ and \NovA. Bound halos are identified using AHF (see \S\ref{sec:wdmics}); these are overplotted as red filled circles, with a size proportional to their virial radii. In the \Ramses\ simulation (left panels), the filaments break up into many small halos. These `spurious' halos have already been shown to be numerical artefacts \citep[e.g.,][]{2007MNRAS.380...93W} and it is encouraging that they are gone in \NovA\ (bottom panels). Once these structures are removed, \NovA\ does a good job of capturing the caustics, fine shells, and criss-crossing filaments that surround galaxies. 

Where caustic and shell structures overlap, AHF often identifies `halos', yet it is not clear if such structures are real or simply transient. We call these `caustic' halos. Their existence -- if real -- is important as it implies that halos can indeed form below the free-streaming mass $M_{\rm fs}$. A detailed exploration of this requires improving the halo finder and studying these structures carefully as a function of time. We will explore this in detail in a forthcoming publication.

\subsubsection{The halo mass function and convergence}\label{sec:massfunc}

\begin{figure}
\includegraphics[width=0.49\textwidth]{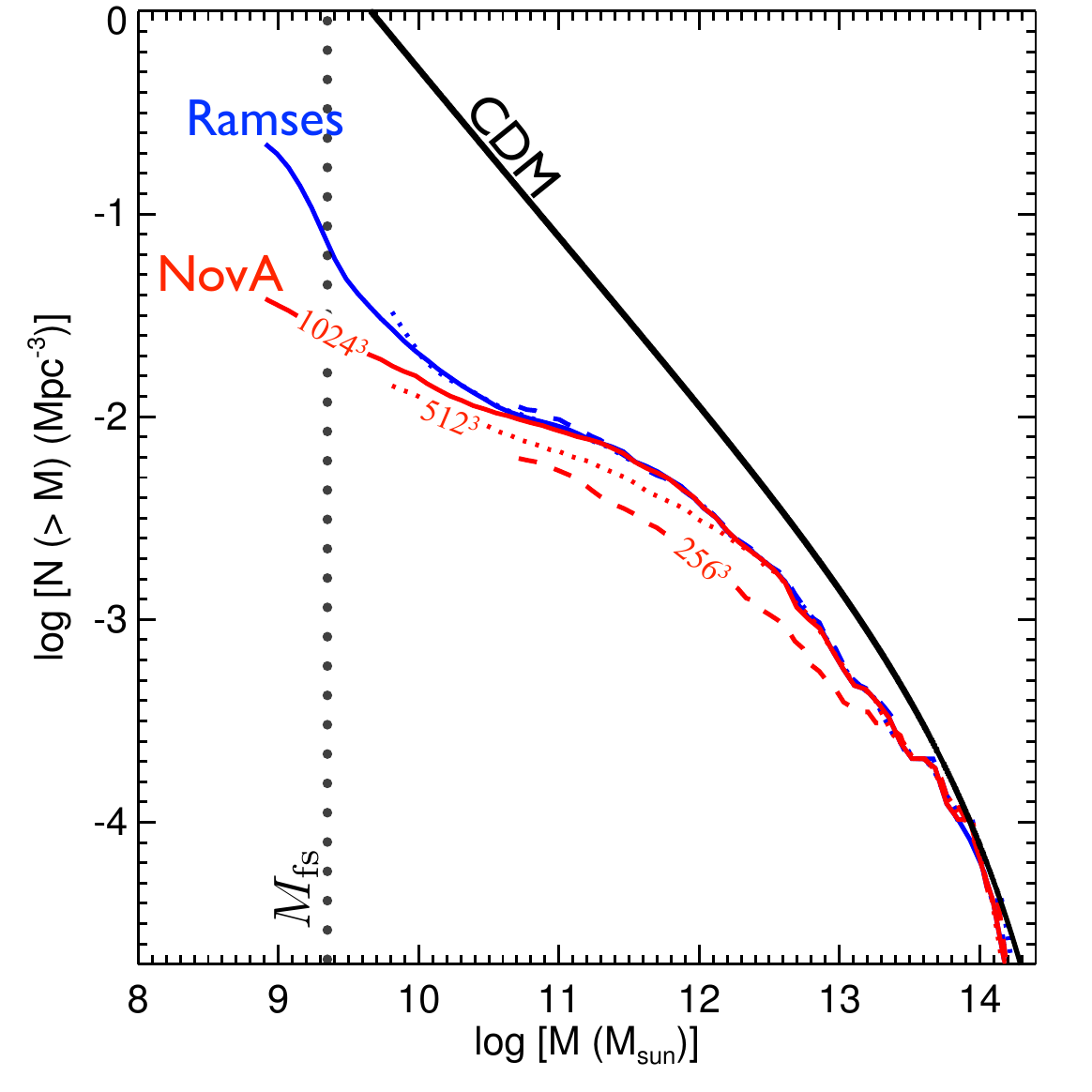}
\caption{Dark matter cumulative halo mass functions in \Ramses\ (blue) and \NovA\ (red) for a 0.2\,keV WDM simulation. In both cases, three different resolutions are marked: $N = 256^3$ (dashed); $N = 512^3$ (dotted); and $N = 1024^3$ (solid) particles. Overlaid is the curve expected for the equivalent cold dark matter simulation (black). Notice also that the standard \Ramses\ simulations converge very slowly with increasing resolution, showing a characteristic upturn at low mass that only strengthens as the resolution increases. By contrast, \NovA\ converges rapidly `from below'. At $N = 1024^3$ particles, the number of halos at low mass in \NovA\ is suppressed with respect to standard \Ramses\ by nearly an order of magnitude. Finally, notice that even in \NovA, the cumulative mass function does not reach a plateau below the free-streaming mass scale: there are significant numbers of halos below $M_{\rm fs}$ (vertical dotted line; see \S\ref{sec:wdmics}). See the text for further discussion of this.}
\label{fig:massfunction}
\end{figure}

In Figure \ref{fig:massfunction}, we compare the AHF cumulative halo mass function for \Ramses\ (blue) with \NovA\ (red) at three numerical resolutions: $N = 256^3$ (dashed), $512^3$ (dotted) and $1024^3$ (solid) particles. Overplotted is the cumulative mass function for the equivalent cold dark matter simulation (black), and the free streaming mass $M_{\rm fs}$ at redshift $z=0$ (dotted line). 

The \Ramses\ simulations show very poor numerical convergence, with a prominent upturn that shifts extremely slowly to lower mass as the resolution is increased. It is interesting that our results for mass function convergence in \Ramses\ are somewhat worse than reported in \citet{2007MNRAS.380...93W}. This likely owes to the fact that we use a different halo finder; we find that switching off the `unbind' feature in AHF leads to very different behaviour, illustrating how sensitive results for WDM simulations are to the choice of halo finder and its chosen settings.

By contrast, \NovA\ shows much better numerical convergence. The mass function rises slowly at the low-mass end with resolution, while remaining unchanged at high mass. Below the free streaming scale, we find nearly an order of magnitude fewer halos in \NovA\ than in \Ramses. We do however find a tendency for the lowest resolutions to oversuppress halos at intermediate mass, although this goes away quickly with increasing resolution.

Our chosen halo finder AHF has been extensively tested on CDM simulations \citep{2011MNRAS.415.2293K}, but has not been used on `spurious halo free' WDM simulations before. As discussed in \S\ref{sec:denfield} (and see Figure \ref{fig:density_composite}), for the simulations we present here, AHF almost certainly misidentifies some features as halos. For this reason, we defer a more careful analysis of halos below the free streaming mass $M_{\rm fs}$ to future work. We discuss how our results compare with other recent determinations in the literature in \S\ref{sec:discussion}. 

Note that our \NovA\ algorithm {\it does} lead to increasing refinement with resolution. This is shown in the final column of Table \ref{tab:tableics}, where we list the maximum refinement level reached for all WDM simulations in this paper. The \NovA\ simulations typically reach $\sim 1$ refinement level less than standard \Ramses, but nonetheless they do continue to refine with increasing resolution. We discuss the effect of varying $\eta$ on the maximum refinement reached and on numerical convergence in \S\ref{sec:etancell} and \S\ref{sec:convergence}.

\subsubsection{Halo density profiles}\label{sec:denprof}

\begin{figure*}
\begin{minipage}[b]{.49\textwidth}
\includegraphics[width=1.00\textwidth]{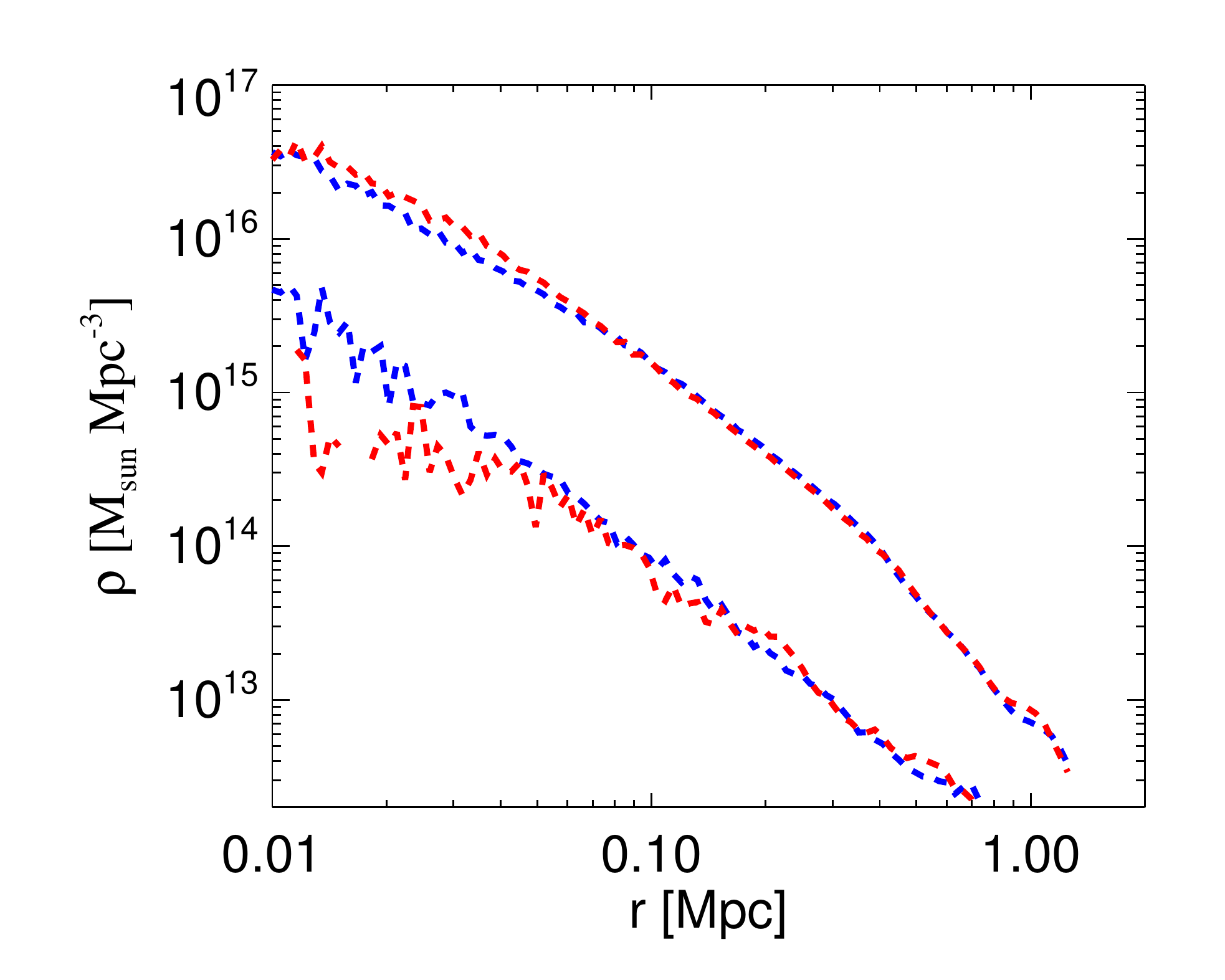}
\end{minipage}
\begin{minipage}[b]{.49\textwidth}
\includegraphics[width=1.00\textwidth]{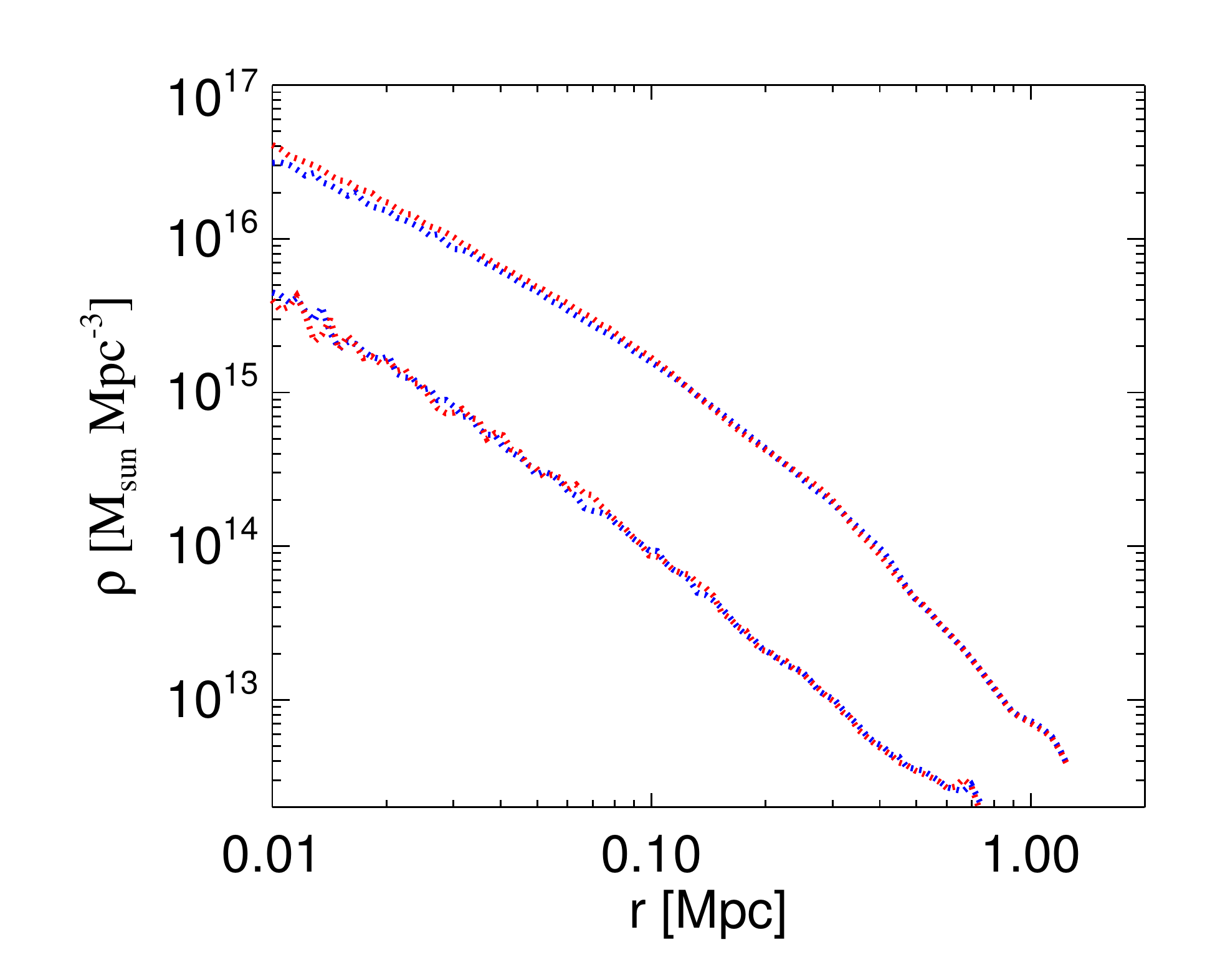}
\end{minipage}
\caption{Dark matter density profiles in a 0.2\,keV WDM simulation for example halos with virial masses $\sim 10^{14} M_{\rm sun}$ (upper lines) and $\sim 10^{12} M_{\rm sun}$ (lower lines) in \Ramses\ (blue) and \NovA\ (red) in the $256^3$ (left) and $512^3$ (right) resolutions. The profiles correspond to the same halos at each resolution in each method.}
\label{fig:haloprofiles}
\end{figure*}

In Figure \ref{fig:haloprofiles}, we show dark matter halo density profiles in \Ramses\ and \NovA\ for the $256^3$ and $512^3$ simulations for both low mass ($\sim 10^{12}$\,M$_\odot$) and high mass ($\sim 10^{14}$\,M$_\odot$) halos. It has already been reported previously in the literature that simply forbidding refinement will reduce spurious two-body effects in collisionless $N$-body simulations \citep[e.g.][]{1997ApJ...479L..79M,1998ApJ...497...38S,2013MNRAS.434.1171H}. However, reducing refinement everywhere also leads to the centres of halos -- where galaxies actually reside -- becoming unresolved. In \NovA, we attempt to obtain the best of both worlds as the algorithm leads to derefinement in highly-anisotropic regions while still refining on halo centres (see Figure \ref{fig:hahn_composite}, bottom right panel). For this reason, the \NovA\ halos typically have a density profile in excellent agreement with the \Ramses\ simulation over all of the resolutions studied here. The lowest mass mass halos in \NovA\ are shallower than their \Ramses\ counterparts reflecting the lower refinement level reached. The effect diminishes with resolution, however, demonstrating that \NovA\ is convergent.

\subsection{Varying $N_c$ and $\eta$: Should some filaments fragment after all?}\label{sec:etancell}

In this section, we study the effect of varying the minimum number of particles in a cell $N_c$ and the anisotropy parameter $\eta$. For this paper, we use throughout $\eta = 1 + 0.5 \sigma_\eta$ as outlined in \S\ref{sec:numerics}. We defer a detailed analysis of $\eta$ -- in particular allowing a time or spatially varying $\eta$ -- to future work. 

\begin{figure*}
\includegraphics[width=0.99\textwidth]{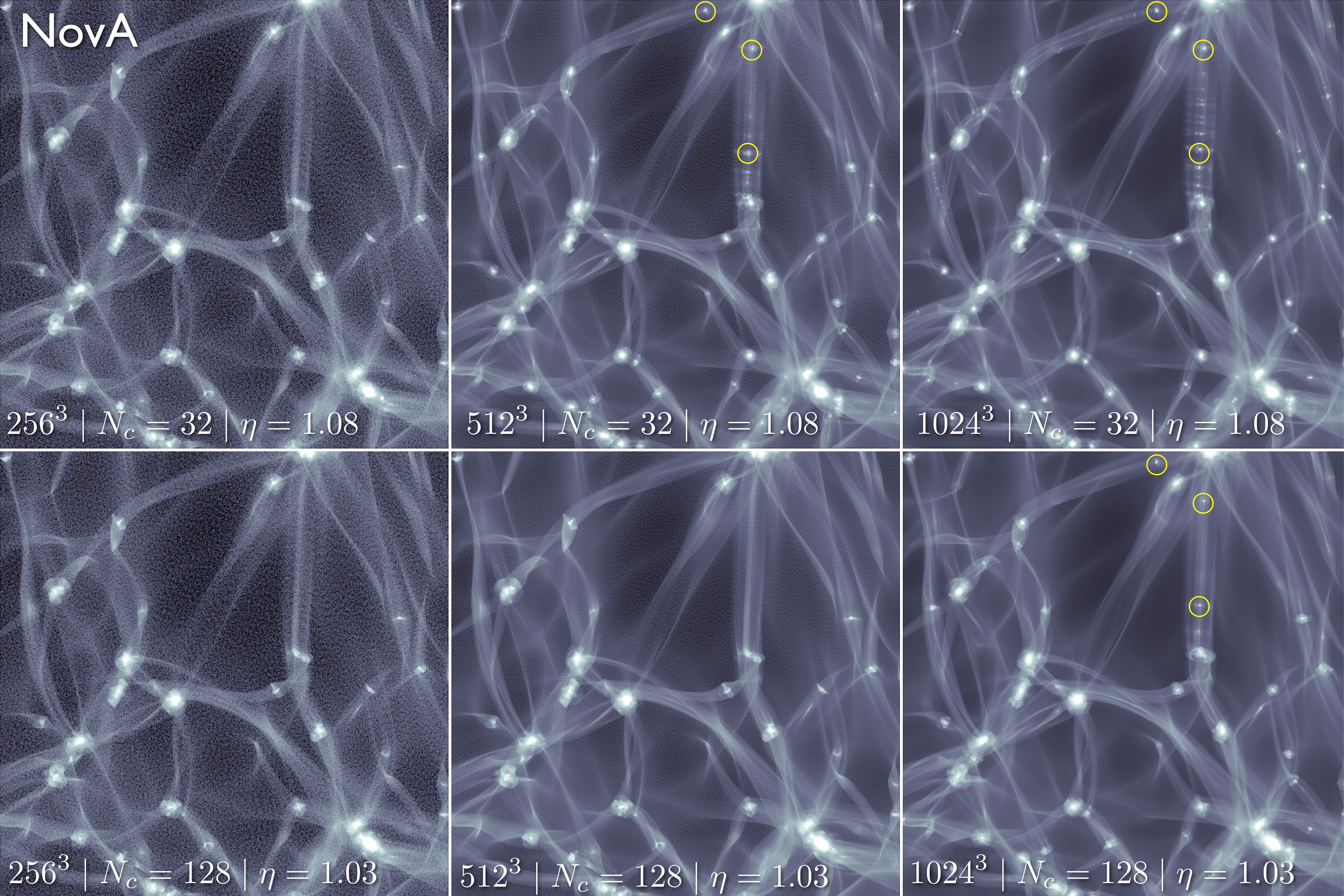}
\caption{Zoom-in on a particularly massive filament (taken from the top middle of the full simulation box; see Figure \ref{fig:density_composite}) for varying $N_c$; $\eta$; and numerical resolution $N$, as marked on each panel. We focus on this particular filament because we find that it is very hard to avoid it fragmenting. At low resolution ($256^3$; left panels), the filament is completely smooth. However, for our default choice of $N_c = 32$, already at $N = 512^3$, the filament begins to fragment (top middle panel), as does another massive filament that connects to the large halo at the edge of the box (see yellow circles). As we raise the numerical resolution $N$ at fixed $N_c$ and $\eta$, the filament fragments further (top right panel). If we raise $N_c$ and lower $\eta$ according to our `noise criteria' (\S\ref{sec:numerics}; bottom panels) then at $N = 512^3$ the filament becomes once again smooth (bottom middle panel). However, even for these \NovA\ parameters, raising the resolution to $N = 1024^3$ results in the filament fragmenting (bottom right panel). Interestingly, the largest structures that form in filaments appear in the same place regardless of our choice of \NovA\ parameters; these are marked by the yellow circles.}
\label{fig:filament_composite}
\end{figure*}

In Figure \ref{fig:filament_composite}, we show a zoom-in on a particularly massive filament that can be seen in the top middle of the full simulation box shown in Figure \ref{fig:density_composite}. We focus on this particular filament because we find that it is very hard to avoid it fragmenting. At low resolution ($256^3$; left panels), the filament is completely smooth. However, for our default choice of $N_c = 32$, already at $N = 512^3$, the filament begins to fragment (top middle panel), as does another massive filament that connects to the large halo at the edge of the box (see yellow circles). As we raise the numerical resolution $N$ at fixed $N_c$ and $\eta$, the filament fragments further (top right panel). If we raise $N_c$ and lower $\eta$ according to our `noise criteria' (\S\ref{sec:numerics}; bottom panels) then at $N = 512^3$ the filament becomes once again smooth (bottom middle panel). However, even for these \NovA\ parameters, raising the resolution to $N = 1024^3$ results in the filament fragmenting (bottom right panel). Indeed, the filament breaks up into structure in {\it any} version of \NovA\ where we permit refinement within the filament. Interestingly, the largest structures that form both in this filament and the one to its top left appear always in the same place regardless of our choice of \NovA\ parameters. These structures are marked by the yellow circles (yellow is chosen to avoid confusion with AHF halos in previous plots that are marked in red). The fact that such structures are challenging to avoid and yet also appear always at the same locations suggests that they may be physically correct. Such behaviour is certainly rather different from the regularly spaced fragments that form along filaments in the standard \Ramses\ simulations (Figure \ref{fig:density_composite}). We discuss this further in \S\ref{sec:discussion}.

\section{Discussion}\label{sec:discussion} 

\subsection{Numerical Convergence in \NovA}\label{sec:convergence}

It is challenging to quantitatively test convergence in \NovA\ because of uncertainties in the halo finding. For example, \citet{2007MNRAS.380...93W} using Friends-of-Friends (FoF) report a definite, if slow, shift of the upturn in the halo mass function to lower mass with increasing resolution. Using AHF, we find also a shift in the upturn but it is substantially slower (see Figure \ref{fig:massfunction}). Taken at face value, it is not at all clear that \Ramses\ will converge on the correct solution with increasing $N$ -- at least not when using AHF.

By contrast, convergence in \NovA\ seems healthier. At low mass, the mass function is suppressed causing a `convergence from below' with increasing $N$. At low resolution, there is a clear over-suppression in halos even at high mass (at $N=256^3$ \Ramses\ outperforms \NovA). But with increasing $N$, the situation rapidly improves. We find a gently rising mass function with few low mass halos, but nonetheless some. Again, owing to difficulties with halo finding, we cannot yet determine whether these low mass halos are real; a fault of the halo finder; or evidence that \NovA\ requires further improvement. We defer a careful analysis of this to a future work where we will improve on the halo finder and study the time evolution of halos.

\subsection{Comparison with recent work in the literature} 

Despite the nearly an order of magnitude suppression in spurious halos in \NovA, at first sight the mass function appears to rise substantially more steeply towards low mass than that reported recently in \citet{2013MNRAS.434.3337A}; using their new `folding phase sheet' methodology, they find no halos below $M_{\rm fs}$. However, there are three confounding factors that make a comparison difficult. Firstly, \citet{2013MNRAS.434.3337A} use a `Friends of Friends' (FoF) halo finder similarly to \citet{2007MNRAS.380...93W}, whereas we use AHF (that includes, for example an `unbinding' procedure that discards halos that are not gravitationally self-bound). Secondly, they throw out all halos with overlapping virial radii. This has the effect that all substructure halos are removed, providing a better comparison with semi-analytic theories for structure formation like Press-Schechter \citep{1974ApJ...187..425P}. However, such a method will also remove half of all binary systems. Finally, they further prune their mass function through a visual inspection of halos. This is necessary due to the FoF algorithm picking up many false positives. We find fewer false positives when using AHF, but many of the issues they report with halo finding in WDM resonate with our findings here. As pointed out in \S\ref{sec:denfield} (and see Figure \ref{fig:density_composite}), we do see many suspicious structures identified by AHF as bound halos that are unlikely to host galaxies. Given the difficulty of halo finding in WDM, we defer a more careful analysis of halos below $M_{\rm fs}$ to a future work where we will look critically at the time evolution of halos in \NovA\ and attempt to improve on existing halo finding algorithms that have all been tuned to work well only on CDM simulations \citep[e.g.][]{2011MNRAS.415.2293K}.

\subsection{Filaments that physically fragment?} 

In crafting a scheme that avoids spurious fragmentation of filaments, we should be mindful of not throwing the baby out with the proverbial bathwater. Are we really sure that no filament should fragment, or that all such fragments reported in $N$-body simulations are spurious? Our results in \S\ref{sec:etancell} strongly suggest that at least some filaments may be unstable to physical fragmentation. From a theoretical point of view, this should perhaps not be surprising. It has long been known that infinite self-gravitating cylinders are unstable to fragmentation \citep{1964ApJ...140.1529O,1984sv...bookR....F}. However, cosmological filaments are both finite and {\it expanding}, and given sufficient expansion, self-gravitating cylinders become unconditionally stable \citep{2011MNRAS.415.1569S}. We will explore this further in future work, but the issue is an important one: if filaments are physically unstable then halos can collapse below $M_{\rm fs}$ and isolated small galaxies can be expected to exist even in warm/hot dark matter cosmologies.

\section{Conclusions}\label{sec:conclusions}

We have introduced a {{\sc Nov}}el form of {{\sc A}}daptive softening (\NovA) for collisionless $N$-body simulations, implemented in the \Ramses\ adaptive mesh refinement code. In \Ramses\ -- that we refer to as a `standard $N$-body method' -- cells are only split if they contain more than eight particles (a mass refinement criterion). We introduced an additional criterion that the cell be sufficiently isotropic, as measured by the ratio of the maximum to minimum eigenvalues of its moment of inertia tensor: $\eta = \lambda_{\rm max}/\lambda_{\rm min}$. In this way, collapse is only refined if it occurs along all three axes, ensuring that the softening $\epsilon$ is always of order twice the largest inter-particle spacing in a cell. This more conservative force softening criterion was designed to minimise spurious two-body effects, while maintaining high force resolution in collapsed regions of the flow.

We tested \NovA\ using an antisymmetric perturbed plane wave collapse (`\Hahn' test) before applying it to warm dark matter (WDM) simulations. Our key results are as follows: 

\begin{itemize}
\item We used the \Hahn\ test to show that -- unlike the standard $N$-body method (\Ramses) -- \NovA\ produces no numerical fragmentation while still being able to correctly capture high density features like the fine caustics and shells around the collapsing regions. 

\item For the WDM simulations, we found that \NovA\ converges significantly more rapidly than \Ramses, producing little or no spurious halos on small scales. \NovA\ produces nearly an order of magnitude fewer dark matter halos at low mass as compared to \Ramses, while still being able to correctly resolve high density regions at the centres of massive halos.

\item Despite the large reduction in low mass halos in \NovA, we still found halos below the free streaming mass scale $M_{\rm fs}$. Furthermore, these halos increase in number (albeit slowly) as we increase the numerical resolution. Some of these likely owe to our halo finder (AHF) incorrectly labelling caustics and criss-crossing filaments as halos. Others form as larger halos that form above $M_{\rm fs}$ are tidally stripped. However, some isolated low-mass structures appear to be real. Due to the difficultly of accurate halo identification in WDM, we defer a quantitative analysis of halos below $M_{\rm fs}$ to a forthcoming publication.

\item We highlighted two particularly massive filaments that fragment in any version of \NovA\ where refinement is allowed. Since the most massive fragments appear always at the same locations, we argue that these may be physical. We noted that infinite self-gravitating cylinders are unstable to collapse and so particularly massive cosmological filaments may be physically unstable. We will explore this further in future work, but the issue is an important one: if filaments are physically unstable then halos can collapse below $M_{\rm fs}$ and isolated small galaxies can be expected to exist even in warm/hot dark matter cosmologies.

\end{itemize}
We will use \NovA\ in forthcoming papers to study the issue of halo formation below $M_{\rm fs}$; filament stability; and to obtain new constraints on the temperature of dark matter. 

\section{Acknowledgments}
JIR would like to acknowledge support from SNF grant PP00P2\_128540/1. We would like to thank Oliver Hahn, Aurel Schneider, Romain Teyssier and Joachim Stadel for useful discussions. This research used the {\sc M\"ONCH} supercomputer owned by ETH Z\"urich, and the {\sc Surrey Galaxy Factory} machine at the University of Surrey.

\bibliographystyle{mnras}

\bibliography{references}

\appendix

\section{Why conservative adaptive softening does not solve the problem of spurious halos}\label{conservative-soft}

One of our earlier ideas for solving the spurious halo problem was to use {\it conservative} adaptive force softening as originally suggested by \citet{Price2007} and implemented recently in the \Gadget\ code \citep{2005MNRAS.364.1105S} by \citet{2011MNRAS.417.2846I}. In this Appendix, for completeness, we report the results of this experiment and explain why it failed to produce the desired effect. In fact, such conservative corrections to adaptive force softening make the problem of spurious halos even worse! 

As detailed in \citet{Price2007}, if the force softening varies in space and time then we can still construct a fully conservative $N$-body method by deriving the equations of motion from a discretised Hamiltonian. This results in an additional corrective force to the usual $N$-body equation of motion: 
\begin{equation}
\frac{d {\bf v}_{i,c}}{dt} = -\frac{G}{2}\sum_j m_j \left[\frac{\xi_i}{\Omega_i}\frac{\partial W_{ij}(h_i)}{\partial {\bf r}_{i}}+
\frac{\xi_j}{\Omega_j}\frac{\partial W_{ij}(h_j)}{\partial {\bf r}_{j}}\right]
\label{eqn:sphconsforcecorrect}
\end{equation}
where: 
\begin{equation} 
\Omega_i \equiv \left[1 - \frac{\partial h_i}{\partial \rho_i} \sum_j m_j \frac{\partial W_{ij}(h_i)}{\partial h_i}\right]
\end{equation}
and:
\begin{equation} 
\xi_i \equiv \frac{\partial h_i}{\partial \rho_i} \sum_j \frac{\partial \phi_{ij}(h_i)}{\partial h_i}
\end{equation}
where $G$ is Newton's gravitational constant; $m_j$ is the mass of particle $j$; $W_{ij}$ is a positive definite spherical smoothing kernel; $h_j$ is the smoothing length of particle $j$, here equated with the softening $h_j = \epsilon_j$; ${\bf r}_j$ is the position of particle $j$; and $\phi_{ij} \equiv \phi(|{\bf r}_i - {\bf r}_j|)$ is related to the gravitational potential between particle pairs \citep{Price2007}.

The key thing to note from equation \ref{eqn:sphconsforcecorrect} is that, since the kernel is positive definite, $\frac{\partial W_{ij}(h_i)}{\partial h_i}$ and similar terms are {\it negative definite}; the correction terms $\Omega_i$ are always of order unity; while the $\xi_i$ terms are also {\it negative definite}. This means that the force correction in equation \ref{eqn:sphconsforcecorrect} is negative definite and will lead always to an {\it increased} gravitational force. This force will point to leading order along the density gradient: 

\begin{equation} 
\nabla_i \rho_i = \sum_j m_j \nabla_i W_{ij}(h_i)
\end{equation} 
This is the trouble with the conservative correction terms. When the flow becomes highly anisotropic, the correction terms will act to {\it increase} the force along the short axis leading to even more artificial clumping (see Figure \ref{fig:anisotropycartoon}).

\begin{figure}
\includegraphics[width=0.49\textwidth]{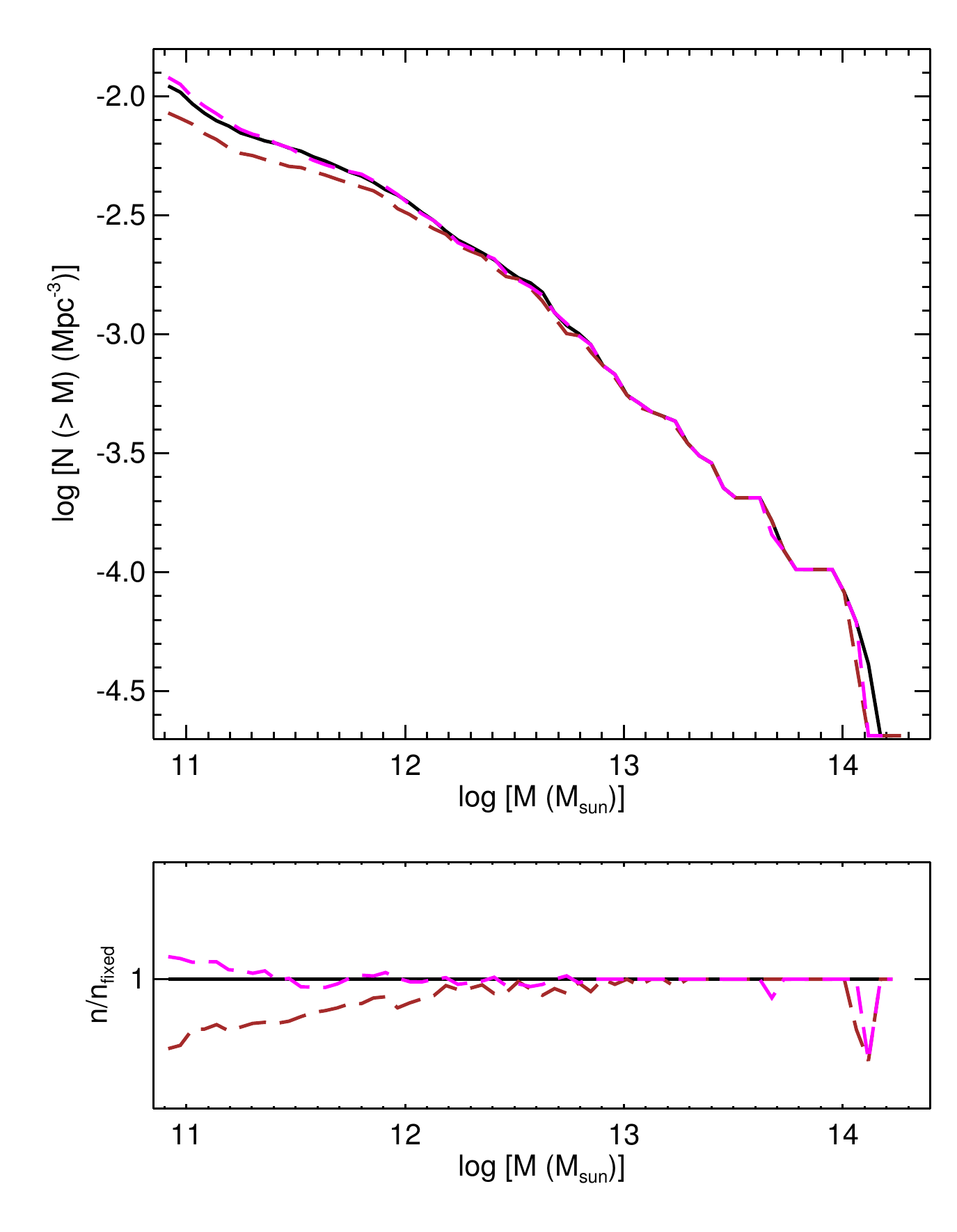}
\caption{Cumulative mass function for \Gadget\ adaptive softening runs, with (magenta) and without (brown) the conservative correction term, compared to the fixed softening reference run (black). Notice that the upturn is more pronounced when the correction term is used. These runs used $N = 128^3$ particles.}
\label{fig:DPornot_mf}
\end{figure}

In Figure \ref{fig:DPornot_mf}, we show the cumulative mass function for a $50$Mpc/h 0.2\,keV WDM simulation run using the standard \Gadget\ $N$-body code (black); with adaptive force softening (brown); and with conservative adaptive force softening (magenta). For this comparison, we used a kernel neighbour number of $N_{\rm neigh} = 60$ since this was shown to be optimal in \cite{2011MNRAS.417.2846I}. Notice that the sharp upturn in the mass function (present in all simulations) is more prominent in the simulation with the conservative correction term (magenta) than even in the fixed softening case (black).

\end{document}